\documentclass[aps, prb, showpacs, twocolumn, 
amssymb,superscriptaddress]{revtex4}

\usepackage{graphicx, bm, amsmath, amsfonts}

\begin{document}


\title{
Entanglement Spectra of the 2D AKLT Model: VBS/CFT Correspondence
}
\author{Jie Lou}
\affiliation{
Institute for Solid State Physics, University of Tokyo,
Kashiwa, Chiba 277-8581, Japan\\
}

\author{Shu Tanaka}
\affiliation{
Department of Chemistry, University of Tokyo, 
Bunkyo-ku, Tokyo 113-0033, Japan\\
}

\author{Hosho Katsura}
\affiliation{
Department of Physics, Gakushuin University, 
Toshima-ku, Tokyo 171-8588, Japan\\
}

\author{Naoki Kawashima}
\affiliation{
Institute for Solid State Physics, University of Tokyo, 
Kashiwa, Chiba 277-8581, Japan\\
}

\date{\today}
\begin{abstract}
We investigate the entanglement properties of the valence-bond-solid (VBS) state 
defined on two-dimensional 
lattices, which is the exact ground state of the Affleck-Kennedy-Lieb-Tasaki model. 
It is shown that the entanglement entropy obeys 
an area law and the non-universal prefactor of the leading term is strictly less than $\ln 2$. 
The analysis of entanglement spectra for various lattices reveals 
that the reduced density matrix associated with the VBS state is closely related to 
a thermal density matrix of a {\it holographic} spin chain, whose spectrum is reminiscent 
of that of the spin-1/2 Heisenberg chain. This correspondence is further supported by 
comparing the entanglement entropy in the holographic spin chain with conformal field theory predictions. 
\end{abstract}


\maketitle
\section{Introduction}
There has been considerable recent interest in understanding 
quantum entanglement in many-body systems.\cite{Amico_review, Eisert_review} 
Entanglement (von Neumann) entropy and the family of R\'enyi entropies are often used to quantify 
the degree of entanglement of a bipartite system consisting of two subsystems $A$ and $B$. 
However, as pointed out by Li and Haldane,\cite{Li_Haldane} 
the entanglement spectrum (ES), the eigenvalue spectrum of the reduced density matrix (RDM) of either system $A$ or $B$, 
provides a more complete description of the ground state and the low-energy states of the system. 
In general, the RDM for $A$ may be written 
as $\rho_{A}=\exp(-H_{\rm E})$ 
with the entanglement Hamiltonian $H_{\rm E}$. 
Li and Haldane demonstrated that the spectrum of $H_{\rm E}$ for fractional quantum Hall states on a sphere 
reflects the gapless edge excitations in the disk geometry. 
Since then, the ES has been applied to other systems such as 
topologically ordered systems~\cite{Lauchli_ES, Thomale_ES, Fidkowski_ES, Qi_Katsura_Ludwig} 
and quantum spin models,\cite{Calabrese_ES, Pollmann_ES, Poilblanc_ES, Yao_Qi_ES} 
and the entanglement gap was proposed as a non-local order in gapless spin chains.\cite{Thomale_spin_chain}

Although the analysis of the ES has turned out to be a powerful tool to extract universal properties, 
the application has so far been limited to one-dimensional (1D) or topological systems. 
To fill this gap in the literature, we examine the ES in the two-dimensional (2D) quantum spin model 
proposed by Affleck, Kennedy, Lieb, and Tasaki (AKLT).\cite{AKLT} 
The AKLT model in one dimension was first introduced as a solvable 
model to illustrate the Haldane-gap phase 
in integer spin chains. The exact ground state is known as the valence-bond-solid (VBS) state. 
The construction of the VBS state generalizes to two and higher dimensions.\cite{KLT, Kirillov_Korepin}  
The VBS state has recently attracted renewed interest from the viewpoint of quantum computation. 
It was proposed that VBS states on a 2D hexagonal lattice can serve as resources for 
measurement-based quantum computation.\cite{Wei_Affleck, Miyake} 
Thus, the study of the ES in VBS states on various 2D lattices 
can shed some light on this important issue.

In this paper, we study numerically the ES in the 2D AKLT model defined on 
various lattices. 
We develop an efficient way to evaluate the RDM using a Monte Carlo (MC) method. 
The ES is then obtained by numerical diagonalization. 
It is shown that the ES is not {\it flat} in 2D, which is in contrast to the 1D case 
where all eigenvalues of the RDM become degenerate in the thermodynamic limit.
\cite{Xu_Katsura_Korepin_Hirano, Korepin_Xu} 
When a system is on a cylinder, we can study the ES as a function of momentum. 
The low-lying spectrum of the entanglement Hamiltonian $H_{\rm E}$ for square (hexagonal) VBS state 
is reminiscent of that of the 1D antiferromagnetic (ferromagnetic) Heisenberg chain. 
This indicates that the entanglement Hamiltonian $H_{\rm E}$ is very close to the Heisenberg Hamiltonian in {\it one dimension}. 
To further support this idea, we introduce a new concept, nested entanglement entropy (NEE), 
and find that, for the square lattice VBS state,  
the low-energy effective Hamiltonian of $H_{\rm E}$ is well described by $c=1$ conformal field theory (CFT) 
despite that the starting quantum state (the AKLT state in this paper) is not critical. 
For a hybrid system consisting of squares and hexagons, we find that the ES is {\it gapped} and 
$H_{\rm E}$ is close to the alternating Heisenberg chain in which ferromagnetic and antiferromagnetic 
interactions are mixed. 

The organization of the rest of the paper is as follows. 
In Sec. II, we first review the construction of the VBS ground states on two-dimensional lattices. 
We then show how to obtain the spectrum of RDM from the Schmidt decomposition of the VBS state. 
We see that the Schmidt coefficients are related to the eigenvalues of the overlap matrix 
(see Eq. (\ref{eq:overlap_matrix}) for the precise definition). 
In Sec. III, we show numerical results of entanglement entropy and spectrum of the VBS states 
on both square and hexagonal lattices. We then introduce the NEE and study its scaling properties. 
We also present an example of the gapped entanglement spectrum, which is obtained from 
the VBS state on the hybrid lattice consisting of squares and hexagons. 
We conclude with a summary in Sec. IV. 
Analytical and numerical approaches for obtaining the overlap matrix are, respectively, 
presented in Appendices A and B.

\section{2D VBS states}
\subsection{AKLT Hamiltonian}
\begin{figure}[b]
\centerline{\includegraphics[width=\columnwidth,clip]{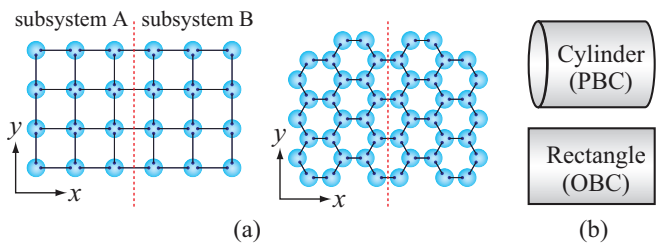}}
\caption{
 (Color online)
 (a) VBS states 
 on the square and hexagonal lattices with reflection symmetry.
 Lattice is separated into two subsystems $A$ and $B$ by the reflection axis 
 (the dotted line). 
 The bulk width is denoted by $L_x$, while 
 the number of bonds on the boundary is $L_y$. 
 Notice that in the 
 hexagonal lattice, $L_x$ is defined as half the number of sites along the $x$-axis. 
 In both cases, $L_x =3$ and $L_y = 4$.
 (b) Boundary conditions. Upper and lower panels show cylindrical (
 PBC) and rectangular(OBC) geometries, 
 respectively.
}
\label{illustration}
\end{figure}
We study the AKLT model 
on the square and hexagonal lattices shown in Fig.~\ref{illustration} (a). 
For simplicity, we focus on the {\it basic} model in which 
there is a single valence bond on each edge. 
The spin-$S$ Hamiltonian for this model is written as a sum of projection operators as
\begin{equation}
H = \sum_{\langle i,j \rangle} A(i,j) \pi_{2S} (i,j),
\end{equation}
where $A(i,j)$ is an arbitrary positive number and $\pi_{2S} (i,j)$ projects 
the total spin of each pair of neighboring spins onto the subspace of $J=2S$. 
The spin magnitude $S$ and the coordination number $z$ are related through $z=2S$,
where $z=4$ for the square lattice, while $z=3$ for the hexagonal lattice. 
The unique ground state (VBS state) of $H$ is written as
\begin{eqnarray}
|\Psi \rangle = \prod_{\langle i,j \rangle} (a^{\dagger}_i b^{\dagger}_j - b^{\dagger}_i a^{\dagger}_j ) |{\rm vac}\rangle,
\label{eq:VBS} 
\end{eqnarray}
where $a^\dagger_i$ and $b^\dagger_i$ are Schwinger bosons at site $i$, 
satisfying 
\begin{equation}
[a_i, a^\dagger_j ] = [b_i, b^\dagger_j ] = \delta_{i,j}
\end{equation}
with all other commutators vanishing,
and $|{\rm vac}\rangle$ denotes the vacuum state.\cite{Arovas_Auerbach_Haldane} 
With the constraint that the total number of bosons at each site is $2S$, 
the original spin operators can be represented by the Schwinger bosons as 
\begin{equation}
S^+_i = a^\dagger_i b_i,~~S^-_i = b^\dagger_i a_i,~~S^z_i = \frac{1}{2}(a^\dagger_i a_i - b^\dagger_i b_i). 
\end{equation}
We will study VBS states defined on lattices with both 
periodic (PBC; cylindrical geometry) and open (OBC; rectangular geometry)
boundary conditions\cite{footnote:boundary} shown in Fig. \ref{illustration} (b). 

\subsection{Reduced density matrix}
Let us consider a bipartition 
into two subsystems $A$ and $B$, each of which is a mirror image of the other, 
and the boundary between them is chosen to be the reflection axis. 
We denote the bulk width of $A$ ($B$) by $L_x$ and the number of bonds on the boundary by $L_y$. 
The Hamiltonian can be written in the form 
$
H = H_{A} + H_{B} + H_{AB},
$
where $H_{A}$ and $H_{B}$ denote the Hamiltonians within subsystems $A$ and $B$, respectively, while $H_{AB}$ contains all the interactions across the boundary. 
The RDM for $A$ describing the entanglement between the two subsystems is defined by
$\rho_{A} = {\rm Tr}_{B} [|\Psi \rangle \langle \Psi|]/{\cal N}$, 
where ${\cal N}= \langle \Psi| \Psi \rangle$. 
To obtain the spectrum of $\rho_{A}$, we follow the same approach used 
in Ref. \onlinecite{KKKKT}. 
The ground state of the system can be written as 
\begin{eqnarray}
|\Psi\rangle = \sum_{\alpha} |\phi^{[{A}]}_{\alpha}\rangle \otimes |\phi^{[{B}]}_{\alpha}\rangle,
\end{eqnarray}
where $\alpha=1,2,...,2^{L_y}$ and 
$\{ |\phi^{[{A}]}_{\alpha}\rangle \}$ ($\{ |\phi^{[{B}]}_{\alpha}\rangle \}$)  
is a set of degenerate ground states of $H_A$ ($H_B$), each of which is characterized 
by a configuration of bosons at the boundary sites. 
These states are linearly independent but not orthogonal. 
We now introduce overlap matrices as 
\begin{eqnarray}
M^{[{A}]}_{\alpha\beta}=\langle\phi^{[{A}]}_{\alpha}|\phi^{[{A}]}_{\beta}\rangle,
\quad
M^{[{B}]}_{\alpha\beta}=\langle\phi^{[{B}]}_{\alpha}|\phi^{[{B}]}_{\beta}\rangle
\label{eq:overlap_matrix}
\end{eqnarray} 
for two subsystems, respectively.
Due to the reflection symmetry, 
one can show that 
$M^{[{A}]} = M^{[{B}]} = M$ 
and $M$ is real symmetric and positive definite. 
One can then construct the orthonormal states out of $\{ |\phi^{[{A}]}_{\alpha}\rangle \}$
($\{ |\phi^{[{B}]}_{\alpha}\rangle \}$) 
using an orthogonal matrix $O$ that diagonalizes $M$, i.e., $O^{\rm T}M O ={\rm diag}(d_1,d_2,...,d_{2^{L_y}})$:
\begin{eqnarray}
|\psi^{[A]}_\alpha \rangle = \frac{1}{\sqrt{d_\alpha}} \sum_{\beta} O_{\beta\alpha} |\phi^{[A]}_\beta \rangle, 
\nonumber \\
|\psi^{[B]}_\alpha \rangle = \frac{1}{\sqrt{d_\alpha}} \sum_\beta O_{\beta\alpha} |\phi^{[B]}_\beta \rangle. 
\end{eqnarray}
The ground state of the full system in the new basis is written in the Schmidt decomposition as
\begin{equation}
|\Psi \rangle = \sum_\alpha d_\alpha |\psi^{[A]}_\alpha \rangle \otimes |\psi^{[B]}_\alpha\rangle, 
\end{equation}
and hence $\rho_A = \sum_\alpha d^2_\alpha |\psi^{[A]}_\alpha \rangle \langle \psi^{[A]}_\alpha|/{\cal N}$. 
Therefore, 
the spectrum of $\rho_A$ is equivalent to that of the following matrix
\begin{equation}
{\hat \rho}_A = \frac{M^2}{{\rm Tr}[M^2]},
\label{eq:RDM1}
\end{equation}
where ${\rm Tr}$ denotes the standard matrix trace (not to be confused with ${\rm Tr}_{A}$). 
One can regard ${\hat \rho}_A$ as a thermal density matrix of an auxiliary spin chain, 
which we call a {\it holographic} spin chain. 
The notion of this hidden spin chain was first introduced in Ref.~\onlinecite{KKKKT}. 
The entanglement Hamiltonian is defined via 
${\hat \rho}_A=\exp(-H_{\rm E})$. 

We can obtain the overlap matrix $M$ analytically for VBS states defined on vertical ladders (i.e. $L_x=1$) 
(see Appendix \ref{app:analytical}). 
For VBS states 
on lattices with a larger width $L_x$ ($>1$), 
in which case results converge to the 2D limit, 
we develop an efficient algorithm of MC method 
(see Appendix \ref{app:MC} for details). 
We sample all possible configurations of bosons 
stochastically, and 
each matrix element 
$M_{\alpha\beta}$ can be numerically obtained as the accumulation number of states. 

\section{Entanglement entropy, spectrum, and nested entanglement entropy}
In this section, we show the results of entanglement entropy and spectrum 
for the VBS states obtained by the MC method combined with exact diagonalization. 
As we will show, the ES of the square (hexagonal) VBS state resembles 
the energy spectrum of the antiferromagnetic (ferromagnetic) Heisenberg Hamiltonian in {\it one dimension}. 
We also introduce a new quantity, {\it nested entanglement entropy}, and 
further elucidate the relationship between the holographic spin chain and 
the Hamiltonian for the Heisenberg chain. 

\subsection{Entanglement entropy}
Let us first study the entanglement entropy (EE) which 
can be obtained from 
\begin{eqnarray}
{\cal S}=-{\rm Tr}[{\hat \rho}_{A} \ln {\hat \rho}_{A}] =-\sum_{\alpha}p_{\alpha}\ln p_{\alpha}, 
\label{entropy}
\end{eqnarray}
where $p_{\alpha}$ ($\alpha=1,2,...,2^{L_y}$) are the eigenvalues of ${\hat \rho}_{A}$.
\begin{figure}
\centerline{
\includegraphics[width=0.95\columnwidth]{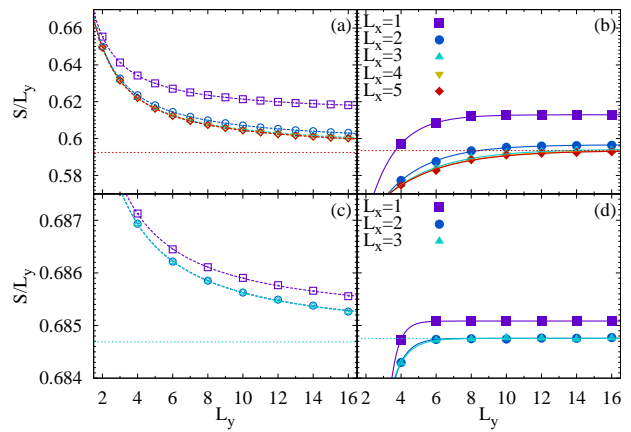}}
\caption{
(Color online)
Entanglement entropy ${\cal S}$ per valence bond across the boundary  
as a function of 
$L_y$ for 
(a) square lattices with rectangular geometry (OBC), 
(b) square lattices with cylindrical geometry (PBC), 
(c) hexagonal lattices with OBC, and 
(d) hexagonal lattices with PBC. 
A further increase of $L_x$ 
does not affect the results much. 
The extrapolations to infinite size are indicated by the broken lines. 
}
\label{EE}
\end{figure}
The obtained EE divided by the boundary length for 
the square and hexagonal lattices 
are shown in Fig.~\ref{EE}.
In both cases, we investigate 
lattices with both PBC and OBC,
results of which are expected to be equivalent in the thermodynamic limit ($L_y \to \infty$).
The EE per unit boundary length ${\cal S}/L_y$ in 2D AKLT model is found to be strictly less than
$\ln 2=0.693147$, which is in contrast to the result for the 1D AKLT model.\cite{Fan_Korepin, Xu_Katsura_Korepin_Hirano, Korepin_Xu} 
Note that the area law is satisfied, as 
${\cal S}/L_y$ approaches to a constant in the thermodynamic limit.

The difference between 1D and 2D EE results can be understood in terms of ``valence bond loops" (VBL), 
which are formed by sequences of singlets 
connected head-to-tail.
(A precise definition of VBL in the context of Schwinger boson can be found 
in Appendix \ref{app:MC}.)
The existence of such loops is a unique feature of 2D VBS states,
which provides additional correlations between two separated singlets. 
As a result, we expect extra correlations between boundary bosons in the 2D VBS states,  
which lead to modification of the EE from the 1D result.
The possibility of forming a valence bond loop decays exponentially fast with respect to its length.
(The reason is discussed in Appendix \ref{app:MC}).
This can be related to multiple observed pheonomena of EE from our numerical calculation.

First of all, in VBS states on the square lattice, the magnitude of deviation from the 1D result (${\cal S}/L_y=\ln 2$)
is significantly larger than that of the hexagonal lattice. 
To understand this, we observe that elemental loops (shortest VBLs that have largest possibility)
connects two boundary sites directly in the square lattice,
while an extra site is in between two boundary spins in the hexagonal lattice.
As a result, the amplitude of elemental loops is higher in the square lattice.
In the end, we obtain considerably more VBLs, which then induce larger amount of extra correlations on the boundary
in the square lattice than in the hexagonal lattice.

Meanwhile, increasing the number of sites $L_x$ in the bulk direction provides additional valence bond loops
which involve sites not only on the boundary, but also inside the two-dimensional lattice. 
Hence the EE is further reduced when expanding bulk width $L_x$. 
The effect converges exponentially fast, however. 
The 2D limit has been almost reached at $L_x=5$ and $L_x=3$ for square and hexagonal VBS states, respectively.
It is again related to the fact that the amplitude of a VBL decreases with its length.

We also observe that the thermodynamic limit ($L_y \to \infty$) of EE is achieved with totally
different behaviors for lattices with rectangular (OBC) and cylindrical (PBC) geometries, 
as shown in Fig.~\ref{EE}.
For VBS states on a lattice with OBC, 
${\cal S}/L_y$ is found to scale with the boundary length ($L_y$) as
\begin{eqnarray}
\frac{\cal S}{L_y} = \sigma + \frac{C_1}{L_y} - \frac{C_2}{L_y \log L_y},
\end{eqnarray}
where $C_1, C_2$ $(>0)$ are the fitting parameters  
and $\sigma$ denotes the EE per boundary length in the thermodynamic limit. 
The magnitude of $C_2$ is significantly less than $C_1$, as shown in Table~\ref{table:fitting}. 
In contrast, for a lattice wrapped on a cylinder (PBC), the thermodynamic limit is approached exponentially fast in $L_y$, 
i.e., $\sigma-{\cal S}/L_y \propto \exp(-L_y/\xi)$, where $\xi$ (also shown in the table) 
is of the order of the lattice spacing. 
\begin{table}[h]
 \begin{center}
  \begin{tabular}{c|ccc|cc}
   \hline\hline
   \multicolumn{6}{c}{square}\\
   \hline
   \multicolumn{1}{c|}{}&
   \multicolumn{3}{|c|}{OBC}&
   \multicolumn{2}{|c}{PBC}\\
   \hline
   & $\sigma$ & $C_1$ & $C_2$ & $\sigma$ & $\xi$ \\
   \hline
   $L_x=1$ & $0.61277(3)$ & $0.0865(3)$ & $0.0009(2)$ & $0.6129(1)$ & $1.54(5)$\\
   $L_x=2$ & $0.59601(4)$ & $0.1127(4)$ & $0.0034(3)$ & $0.5966(2)$& $2.4(1)$\\
   $L_x=3$ & $0.59322(5)$ & $0.1196(5)$ & $0.0049(3)$ & $0.5942(3)$ & $2.9(2)$\\
   $L_x=4$ & $0.59259(5)$ & $0.1223(5)$ & $0.006(3)$ & $0.5936(4)$ & $3.1(2)$\\
   $L_x=5$ & $0.59246(4)$ & $0.1227(5)$ & $0.006(3)$ & $0.5934(4)$ & $3.1(2)$\\
   \hline\hline
  \end{tabular}

  \vspace{5mm}

 \begin{tabular}{c|ccc|cc}
   \hline\hline
   \multicolumn{6}{c}{hexagonal}\\
   \hline
   \multicolumn{1}{c|}{}&
   \multicolumn{3}{|c|}{OBC}&
   \multicolumn{2}{|c}{PBC}\\
   \hline
   & $\sigma$ & $C_1$ & $C_2$ & $\sigma$ & $\xi$ \\
   \hline
  $L_x=1$ & $0.68502(2)$ & $0.0092(6)$ & $0.0011(7)$ & $0.68508(6)$ & $0.4(1)$ \\
  $L_x=2$& $0.68469(3)$ & $0.0098(6)$ & $0.0012(8)$ & $0.684757(4)$ & $0.6(1)$ \\
  $L_x=3$ & $0.68468(3)$ & $0.0098(6)$ & $0.0013(7)$ & $0.684754(5)$ & $0.7(1)$\\
  \hline\hline
 \end{tabular}
  \caption{
  Obtained fitting parameters for entanglement entropy by the method of least squares.
  }
  \label{table:fitting}
 \end{center}
\end{table}

Such distinctively different behaviors of EE for PBC and OBC lattices
are closely related to VBL formations. 
For example, for the square lattice
the leading correction to boundary spin correlations and EE 
is associated with a group of elemental valence bond loops mentioned above, 
each connecting two nearest neighbor boundary spins.
In a lattice with rectangular geometry (OBC), the number of such elemental loops is always
$1$ less than 
the number of boundary sites $L_y$.
As a result, the leading correction to entropy per boundary spin is proportional to
$(L_y-1)/L_y$, which approaches $1$ in the thermodynamic limit.
Accordingly, we expect that the entanglement entropy converges slowly following the function $1/L_y$,
as observed in Fig.~\ref{EE}. 

For a lattice wrapped on a cylinder (PBC), winding loops associated with periodic boundary 
provide extra spin-spin correlations to the system.
The EE in a finite system is lower than the result in the thermodynamic limit,
due to the fact that such winding valence bond loops are more prominent in a smaller lattice
compared with those in a larger one. 
Furthermore, amplitudes associated with winding loops decay exponentially fast according to the 
number of boundary spins $L_y$, which is of the same order of winding loops' length.
As a result, we observe a exponentially converging behavior of the entanglement entropy in a PBC lattice.




\subsection{Entanglement spectrum}
\begin{figure}
\centerline{\includegraphics[width=.9\columnwidth, clip]
{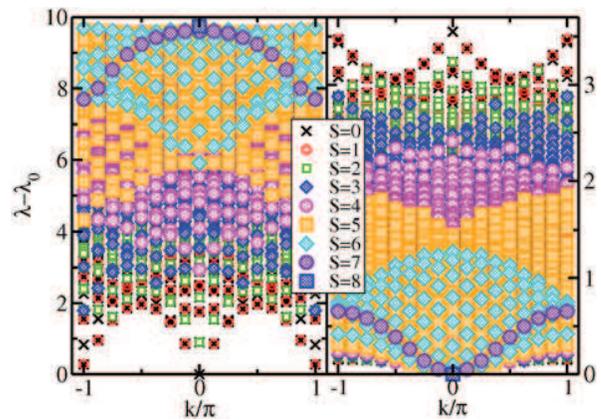}}
\caption{
(Color online) 
Entanglement spectra 
of the (left) square and 
(right) hexagonal VBS states with cylindrical geometry (PBC). 
In both cases, 
$L_y=16$ and 
$L_x=5$, in which case results have converged to the two-dimensional limit.
The ground state energy of $H_{\rm E}$ is denoted by $\lambda_0$. 
The total spins $S$ of the states are marked by different symbols. 
}
\label{ES}
\end{figure}

Let us now consider the ES defined as a set of eigenvalues 
of the entanglement Hamiltonian $H_{\rm E}:=-\ln {\hat \rho}_A$. 
In a system with 
PBC, we are able to study the ES 
as a function of momentum in the $y$-direction ($k$ in Fig.~\ref{ES}). 
As shown in Fig.~\ref{ES}, the ES obtained from 
the square (hexagonal) VBS states
resembles the spectrum of the spin-1/2 antiferromagnetic (ferromagnetic) Heisenberg chain. 
The reason for the dependence of the ES on the lattice structure is as follows: 
Although both the square and hexagonal lattices are bipartite,  
neighboring boundary spins belong to the different sublattices in the case of the square VBS state.
As a result, the entanglement Hamiltonian is reminiscent of the AFM Heisenberg chain. 
In contrast, in the hexagonal lattice, all boundary spins belong to the same sublattice, 
hence the FM spectrum appears. 
The lowest-lying modes in the left panel can be identified as the des Cloizeaux-Pearson 
spectrum in the AFM Heisenberg chain~\cite{desCloizeaux_Pearson}. 
In the right panel, the excitations with the total spin $S=7$ look like the ordinary spin-wave 
spectrum in the FM Heisenberg chain. In fact, due to the translational symmetry in the $y$ direction, 
the Bloch theorem applies and the single-magnon states in the FM chain are {\it exact} eigenstates 
of the RDM for the hexagonal VBS state. 


\subsection{Nested entanglement entropy}
In order to further establish the correspondence between the holographic spin chain 
and the Hamiltonian for the Heisenberg chain, 
we introduce a measure, which we call {\it nested entanglement entropy} (NEE), 
and study its scaling properties. 
One might think that the finite-size scaling analysis of the ground state energy $\lambda_0$ 
is sufficient to reveal the CFT structure in the holographic spin chain. 
However, there is a subtle point here. The spectrum of the RDM can only tell us that 
the entanglement Hamiltonian can be expressed as $H_{\rm E} = \beta_{\rm eff} H_{\rm hol}$ 
with the Hamiltonian for the holographic chain $H_{\rm hol}$. 
There is no unique way to disentangle a fictitious temperature $\beta_{\rm eff}$ from $H_{\rm hol}$. 
Assuming that $H_{\rm hol}$ is gapless and its low-energy dispersion is given by $vk$, 
we can estimate $\beta_{\rm eff} v$ from the slope of the modes at $k=0$ 
in the left panel of Fig.~\ref{ES} (see Table \ref{table:beta_eff}). 
The obtained slope shows saturation at about $L_x=4,5$, which suggests that the fictitious temperature is 
presumably nonzero ($\beta_{\rm eff} < \infty$) in the infinite 2D system. 
As we will see, however, the NEE provides a more clear-cut approach to investigate 
the critical behavior of $H_{\rm hol}$ without any assumption. 
\begin{table}[t]
 \caption{\label{table:beta_eff} Slopes ($\beta_{\rm eff}v$) of modes at $k=0$ for $L_y=16$.}
 \begin{ruledtabular}
  \begin{tabular}{lllll}
   $L_x=1$ & $L_x=2$ & $L_x=3$ & $L_x=4$ & $L_x=5$\\
   $1.69000$ & $2.02221$ & $2.13535$ & $ 2.17006$ & $2.18755$
  \end{tabular}  
 \end{ruledtabular}
\end{table}

\begin{figure}
\centerline{\includegraphics[width=4.25cm,clip]{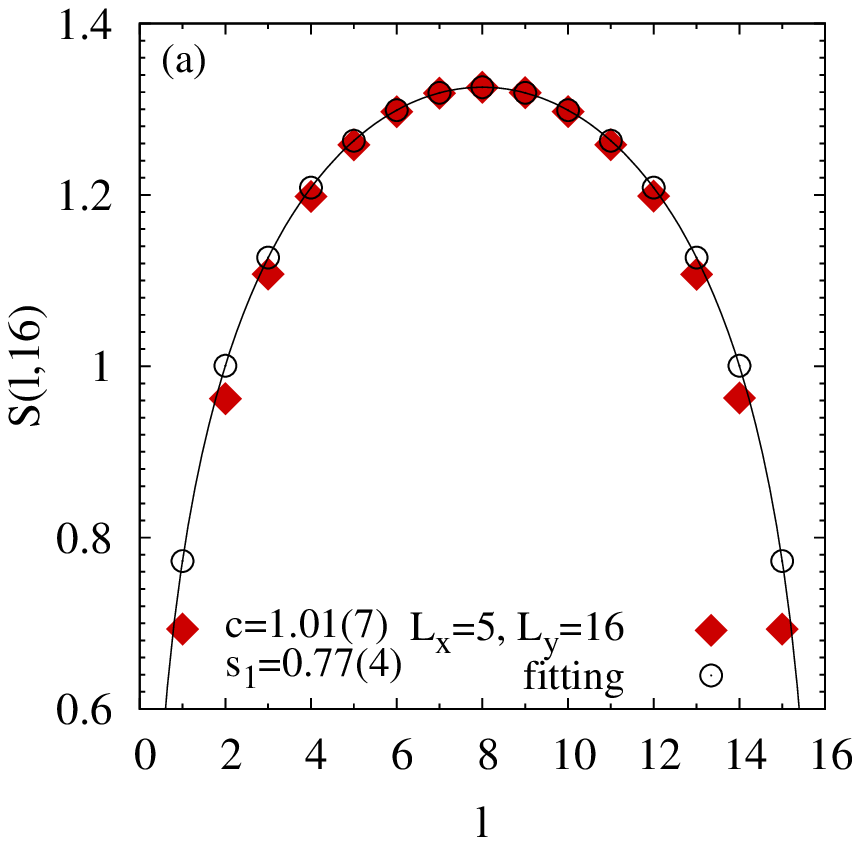}
\includegraphics[width=4.25cm,clip]{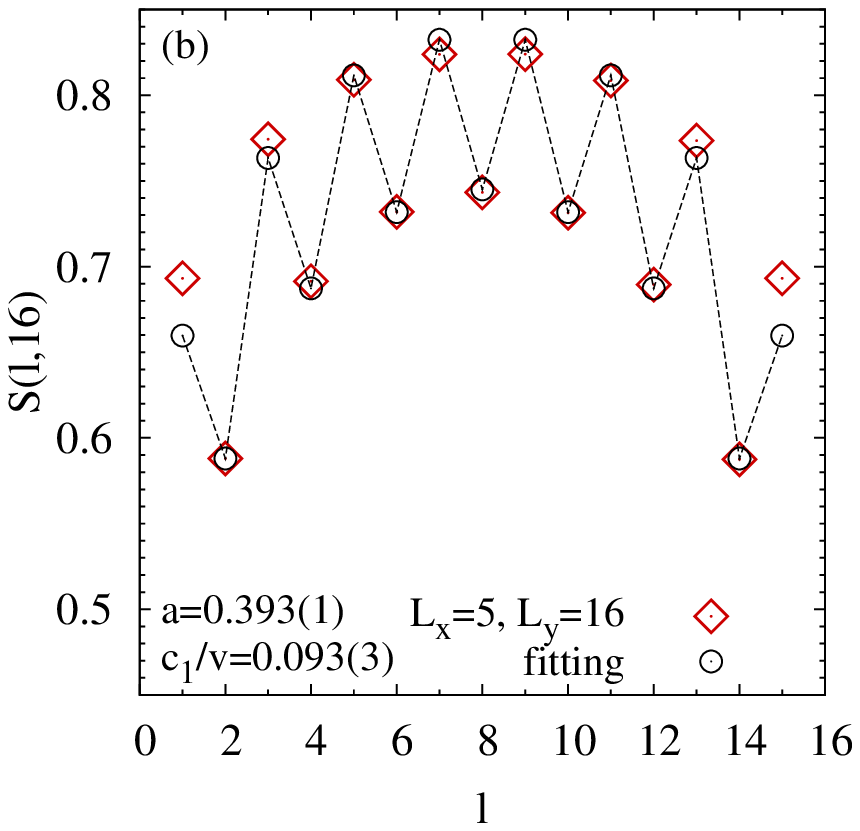}}
\caption{
(Color online)
Nested entanglement entropy $S(\ell, L_y)$ as a function of the sub-chain length $\ell$ for $L_x = 5$ and $L_y = 16$.
(a) and (b) show results obtained for square VBS states 
with PBC and OBC, respectively.
Fits to the CFT predictions, 
Eqs.~(\ref{PBC_function}) and (\ref{OBC_function}), are indicated by open circles. 
The lines are guides to the eye.
}
\label{NEE}
\end{figure}

Let us now give a precise definition of the NEE. 
Based on the ground state of the entanglement Hamiltonian $H_{\rm E}$, 
we construct the nested 
RDM for the sub-chain of length $\ell$ in the holographic spin chain as
\begin{equation}
\rho(\ell)= {\rm Tr}_{\ell+1,\ldots,L_y}[|\psi_0\rangle\langle\psi_0|],
\end{equation}
where $|\psi_0\rangle$ denotes the normalized ground state of $H_{\rm E}$ 
and the trace is taken over the remaining sites 
excluding the sub-chain. 
The NEE 
is then obtained from the nested RDM as
\begin{equation}
{\cal S}(\ell,L_y)=-{\rm Tr}_{1,\ldots,\ell}[\rho(\ell) \ln \rho(\ell)],
\end{equation}
where the trace is over the sites on the sub-chain. 
Since the low-energy physics of the AFM Heisenberg chain is described by $c=1$ CFT, 
the NEE is expected to behave as \cite{Calabrese_Cardy}
\begin{eqnarray}
{\cal S}^{\rm PBC}(\ell,L_y) &=& \frac{c}{3}\ln [f(\ell)] + s_1,
\label{PBC_function} \\
f(\ell) &=& \frac{L_y}{\pi} \sin \left( \frac{\pi \ell}{L_y} \right),
\end{eqnarray}
for a lattice with PBC, where $c$ is the central charge, and $s_1$ is a non-universal constant.
In Fig.~\ref{NEE} (a), we show NEE obtained from the square VBS state 
with $L_x = 5$ and $L_y = 16$. 
The CFT prediction, Eq. (\ref{PBC_function}), well explains the spatial profile of the NEE. 
The fit yields $c=1.01(7)$ which is reasonably close to $c=1$. 

The NEE for $\ell=L_y/2$ (half-NEE) is simplified to be
$
{\cal S}^{\rm PBC} (L_y/2,L_y)= (c/3) \ln (L_y) + {\rm const.}
$
From the data with the finite-size scaling form of the NEE, 
we can also extract the central charge $c$ 
as summarized in Table \ref{table:half_NEE}.  
\begin{table}[t]
 \caption{\label{table:half_NEE} The obtained central charge from half-NEE.}
 \begin{ruledtabular}
  \begin{tabular}{lllll}
   $L_x=1$ & $L_x=2$ & $L_x=3$ & $L_x=4$ & $L_x=5$\\
   $1.007(4)$ & $1.042(4)$ & $1.055(4)$ & $1.056(2)$ & $1.059(2)$
  \end{tabular}  
 \end{ruledtabular}
\end{table}
For the case $L_x=1$, 
$c$ is remarkably close to unity. 
This further supports our conjecture: a correspondence between the entanglement Hamiltonian 
of the 2D AKLT model and the physical Hamiltonian of the 1D Heisenberg chain. 
A slight modification of $c$ is observed as the bulk width increases from $L_x=1$ to $L_x=5$. 

In the case of the rectangular geometry (OBC), a staggered pattern of NEE 
is expected from studies of the standard EE in open spin chains~\cite{Laflorencie_et_al, Hikihara_et_al}: 
\begin{eqnarray}
 {\cal S}^{\rm OBC}(\ell,L_y)&=&\frac{c}{6}\ln [g(2\ell+1)] + a -\frac{\pi c_1}{2v}
\frac{(-1)^\ell}{g(2\ell+1)^{1/2K}},\nonumber  \\
 g(\ell)&=&\frac{2(L_y+1)}{\pi} \sin \left(\frac{\pi \ell}{2(L_y+1)} \right).
\label{OBC_function}
\end{eqnarray}
In Fig.~\ref{NEE} (b), we show the NEE for the square VBS state on the rectangle with $L_x=5$ and $L_y=16$. 
We take the central charge to be $c=1$ and the Tomonaga-Luttinger parameter $K=1$, 
and tune $a$ and $c_1/v$ as fitting parameters. 
The fit is reasonably good except for the boundaries. 
This deviation might be due to logarithmic corrections which appear in the SU(2) symmetric AFM spin chains~\cite{Laflorencie_et_al} 
but are not taken into account in Eq. (\ref{OBC_function}). 

\begin{figure}
\centerline{\includegraphics
[width=0.9 \columnwidth,clip]
{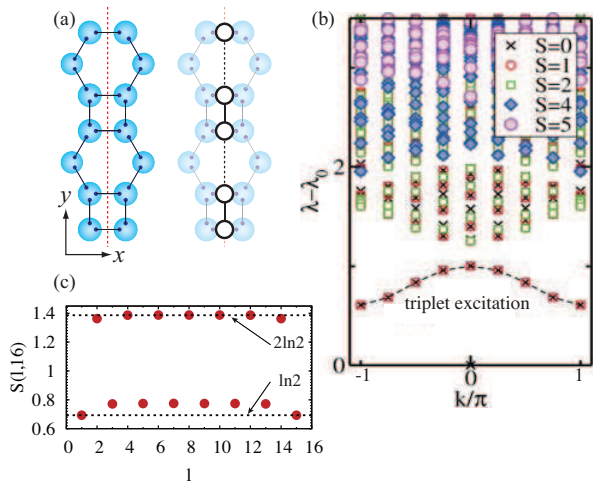}}
\caption{
(Color online) (a) (Left) VBS state on a hybrid lattice.
The red (dotted) line indicates the reflection axis. 
(Right) Corresponding holographic chain. 
The thick and dashed lines indicate AFM and FM bonds, respectively.  
(b) Entanglement spectrum of the VBS state in (a) with $L_y=16$. 
The ground state energy is denoted by $\lambda_0$. 
The dashed curve indicates the spectrum of triplet-pair bound states. 
(c) Nested entanglement entropy ${\cal S}(\ell, L_y)$ as a function of $\ell$ 
for $L_y=16$. 
}
\label{fig5}
\end{figure}
To provide further evidence for the holographic spin chain, we now consider the hybrid system 
comprised of squares and hexagons, shown in Fig.~\ref{fig5}(a). 
It is expected that the entanglement Hamiltonian is described by the FM-AFM alternating Heisenberg model. 
Figure~\ref{fig5}(b) shows the ES for this system. It is clearly seen that there is a gap between 
the ground state and the continuum. The gap rapidly saturates with increasing $L_y$. 
The behavior of the ES is totally consistent with the energy spectrum of 
the alternating Heisenberg chain whose interactions are denoted $J$ and $J'$.\cite{Hida_alternating} 
Comparing the bound-state spectrum in Fig.~\ref{fig5}(b) with that of the triplet-pair excitation in 
the alternating Heisenberg chain,\cite{Hida_alternating}  
we estimate the ratio of the FM exchange divided by the AFM one as $J'/J \sim 0.5$. 
Note that one can further manipulate $J'/J$ of the holographic spin chain by tuning the lattice structure and 
the number of valence bonds on each edge the lattice.  
The NEE associated with the ground state of $H_{\rm E}$ is shown in Fig.~\ref{fig5}(c). 
The NEE is $2\log 2$ when two AFM bonds are cut, whereas it becomes about $\log 2$ when both FM and AFM bonds are cut. 
The obtained result is in good agreement with the standard EE in the alternating Heisenberg chain 
studied in Ref.~\onlinecite{Hirano-Hatsugai}.

\section{Conclusion}
We have studied both the entanglement entropy and spectrum associated with 
the VBS ground state of the AKLT model on various 2D lattices. 
It was shown that the reduced density matrix of a subsystem can be interpreted 
as a thermal density matrix of the holographic spin chain, whose spectrum 
resembles that of the spin-1/2 Heisenberg chain. 
To elucidate this relationship, we have introduced the concept, {\it nested entanglement entropy} (NEE), 
which allows us to clarify this correspondence in a quantitative way without information 
on the fictitious temperatures. 
The finite-size scaling analysis of the NEE revealed that the low-energy physics of the holographic chain 
associated with the square lattice VBS is well described by $c=1$ conformal field theory. 
The NEE was also applied to the hybrid VBS state, the entanglement spectrum of which is gapped, 
and the holographic chain was found to be the alternating Heisenberg chain. 

\section*{Acknowledgment}
The authors thank Toshiya~Hikihara, Vladimir~E.~Korepin, Takafumi~Suzuki, and Hal~Tasaki
for valuable discussions, and Hidetoshi~Nishimori for TITPACK, ver.~2.
This work was supported by Grant-in-Aid for JSPS Fellows (23-7601),
for Young Scientists (B) (23740298), for Scientific Research (B)
(22340111), and for Scientific Research on Priority Areas (19052004).
Numerical calculations were performed on supercomputers 
at the Institute for Solid State Physics, University of Tokyo.

{\it Note added}.---
While completing this work, we became aware of a paper by 
I. J. Cirac et al.~\cite{Cirac_et_al} in which similar results 
were obtained using different approaches. 

\appendix
\section{Analytical method for $L_x=1$}
\label{app:analytical}

In this appendix, we show how to explicitly calculate the reduced density matrix for $L_x=1$ ladders 
with OBC and PBC based on an extension of the method described in Ref. \onlinecite{KKKKT},  
in which the authors analytically obtained entanglement entropies of the VBS states with OBC 
using a transfer matrix technique.
According to Ref. \onlinecite{KKKKT}, the overlap matrix $M$ for vertical ladder, i.e. $L_x=1$ is given by
\begin{widetext}
\begin{eqnarray}
 M = \int 
  \left(
   \prod_{i=1}^{L_y} \frac{(z_i + 1)!}{4\pi} {\rm d}\hat{\Omega}_i
  \right)
  \prod_{j=1}^{L_y}
  \left(
   \frac{1+\hat{\Omega}_j \cdot \vec{\sigma}_j}{2}
  \right)
  \prod_{k=1}^{\tilde{L}_y}
  \left(
   \frac{1-\hat{\Omega}_k \cdot \hat{\Omega}_{k+1}}{2}
  \right),
  \,\,\,
  \tilde{L}_y = 
  \begin{cases}
   L_y - 1 \, &({\rm OBC})\\
   L_y \, &({\rm PBC})
  \end{cases},
\end{eqnarray}
\end{widetext}
where $z_i$ denotes the coordination number at the $i$-th site, $\hat{\Omega}_i$ is the unit vector defined by 
$\hat{\Omega}_i :=(\sin \theta_i \cos \phi_i, \sin\theta_i \sin\phi_i, \cos\theta_i)$, 
and $\vec{\sigma}_j=(\sigma^x_j, \sigma^y_j, \sigma^z_j)$ is the Pauli matrix vector.
For PBC, $\hat{\Omega}_{L_y+1}=\hat{\Omega}_1$. 
The above expression for $M$ can be derived 
using the Schwinger boson representation and coherent state representation. 
It is convenient to introduce $Z$-matrix defined by
\begin{eqnarray}
 Z(L_y;\left\{ \vec{\sigma}_i \right\})
&:=& \int 
  \left(
   \prod_{i=1}^{L_y} \frac{{\rm d}\hat{\Omega}_i}{4\pi}
  \right)
  \prod_{j=1}^{L_y}
  (
   1+\hat{\Omega}_j \cdot \vec{\sigma}_j
  )
  \nonumber \\
  &\times &
  \prod_{k=1}^{\tilde{L}_y}
  (
   1-\hat{\Omega}_k \cdot \hat{\Omega}_{k+1}
  ),
\end{eqnarray}
Since the only difference between $M$ and $Z$ is the overall factor, 
the spectrum of $\hat{\rho}_A$ is equivalent to that of the following matrix:
\begin{eqnarray}
 \rho_A' := 
  \frac{Z(L_y;\left\{ \vec{\sigma}_i\right\})^2}{{\rm Tr}\,[Z(L_y;\left\{ \vec{\sigma}_i\right\})^2]},
\end{eqnarray}
where ${\rm Tr}$ is taken over $\sigma$-spin spaces.
Then in order to calculate entanglement properties of the VBS state on a vertical ladder, 
it is necessary to obtain the explicit form of $Z$-matrix.
There is a useful relation shown in lemma 3.3 in Ref. \onlinecite{KLT}:
\begin{eqnarray}
 \int \frac{{\rm d}\hat{\Omega}_i}{4\pi}
  ( \vec{\sigma}_i \cdot \hat{\Omega}_i )
  ( \hat{\Omega}_i \cdot \vec{\sigma}_j )
  = q \vec{\sigma}_i \cdot \vec{\sigma}_j,
\end{eqnarray}
where $q=1/3$.
Using the above relation we obtain the following relation:
\begin{eqnarray}
 \label{eq:z-matrix-square}
 &Z(L_y;\left\{\vec{\sigma}_i\right\})
  &= \sum_{\left\{ i_1, \cdots, i_{2n}\right\}}
 q^n (-q)^{l_1} \cdots (-q)^{l_n}\nonumber\\
 && \times \left( \vec{\sigma}_{i_1} \cdot \vec{\sigma}_{i_2} \right)
  \cdots 
  \left( \vec{\sigma}_{i_{2n-1}} \cdot \vec{\sigma}_{i_{2n}} \right),
\end{eqnarray}
where $l_j = {\rm mod}(i_{2j}-i_{2j-1},L_y)$ and $1 \le i_1 < i_2 < \cdots < i_{2n-1} \le L_y$.
The summation is taken over admissible combinations of $\{i_j\}_{j=1,...,2n}$ for $0 \le n \le N/2$.
It is useful to express Eq.~(\ref{eq:z-matrix-square}) in terms of graphical representations.

\begin{figure}[b]
 \begin{center}
  \includegraphics[scale=1]{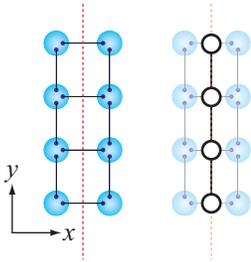}
  \caption{
  (Color online) 
  Relationship between the original VBS state on a square-lattice vertical ladder (left) and 
  a holographic spin chain (right). The circles in the right panel indicate the sites in the spin chain, 
  on which the Pauli vectors $\left\{ \vec{\sigma}_i \right\}$ in Eq. (\ref{eq:z-matrix-square}) act. 
  }
 \label{fig:z-matrix-square}
 \end{center}
\end{figure}

\begin{figure*}[htb]
 \begin{center}
  \includegraphics[scale=0.9]{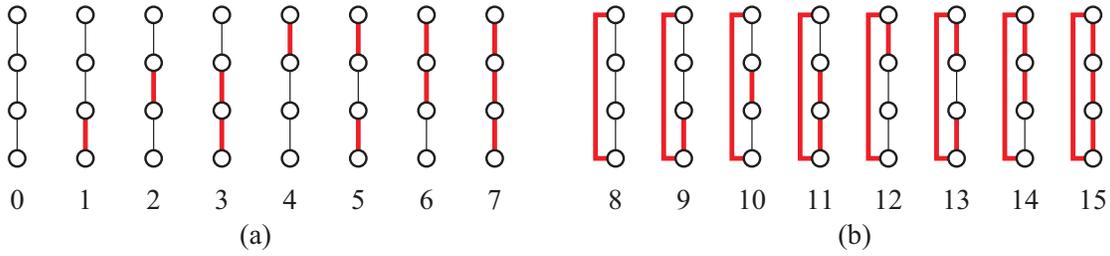}
  \caption{
  (Color online) 
  Graphical representation of all admissible graphs.
  When we study the OBC case, we have to consider only the configurations depicted in (a), whereas we have to consider the configuration depicted in (b) in addition to that depicted in (a), we study the PBC case.
  }
 \label{fig:Ly4}
 \end{center}
\end{figure*}

Figure \ref{fig:z-matrix-square} shows the relationship between the original VBS state and the one-dimensional holographic spin chain.
The circles in the right panel in Fig.~\ref{fig:z-matrix-square} indicate the sites in the spin chain, 
  on which the Pauli vectors $\left\{ \vec{\sigma}_i \right\}$ in Eq. (\ref{eq:z-matrix-square}) act. 
$Z$-matrix is represented by sum of admissible configurations each of which has a corresponding graph (see Fig.~\ref{fig:Ly4}).
Here the number of admissible configurations is $2^{\tilde{L}_y}$.

Let us consider the $L_y = 4$ case for illustrative purposes. 
Figure \ref{fig:Ly4}(a) shows all admissible configurations for OBC using a bit representation in which the thin and thick lines denote $0$ and $1$, respectively. 
The end points of thick lines in Fig.~\ref{fig:Ly4} indicate $\left\{ i_j\right\}$.
For example, the contributions of the states 3 and 5 to the $Z$-matrix are expressed as
\begin{eqnarray}
 &q^1 (-q)^2 \left( \vec{\sigma}_1 \cdot \vec{\sigma}_3 \right),\\
 &q^2(-q)^1(-q)^1 \left( \vec{\sigma}_1 \cdot \vec{\sigma}_2 \right)
  \left( \vec{\sigma}_3 \cdot \vec{\sigma}_4 \right),
\end{eqnarray}
respectively.
Note that $n$ in Eq.~(\ref{eq:z-matrix-square}) corresponds to the number of 
clusters connected by thick lines contiguously 
in each graphical representation. 
Summing all contributions, we arrive at the following expression for the $Z$-matrix:
\begin{eqnarray}
\nonumber 
Z^{\rm OBC}(4;\left\{ \vec{\sigma}_i\right\})
  &&= I - q^2
  \left[ 
   \left( \vec{\sigma}_1 \cdot \vec{\sigma}_2 \right) + 
   \left( \vec{\sigma}_2 \cdot \vec{\sigma}_3 \right) + 
   \left( \vec{\sigma}_3 \cdot \vec{\sigma}_4 \right)   
  \right]\\
  &&+ q^3
  \left[
   \left( \vec{\sigma}_1 \cdot \vec{\sigma}_3 \right) + 
   \left( \vec{\sigma}_2 \cdot \vec{\sigma}_4 \right)
  \right]\nonumber\\
  &&+ q^4
  \left[
   \left( \vec{\sigma}_1 \cdot \vec{\sigma}_2 \right)
   \left( \vec{\sigma}_3 \cdot \vec{\sigma}_4 \right)
   - 
   \left( \vec{\sigma}_1 \cdot \vec{\sigma}_4 \right)
  \right],
\end{eqnarray}
where $I$ is a $2^{L_y}$-dimensional identity matrix.

Next, we examine the case of PBC where the number of admissible configurations is $2^{\tilde{L}_y}(=2^{L_y})$. 
We have to consider the states depicted in Fig.~\ref{fig:Ly4} (b) in addition to the states in Fig.~\ref{fig:Ly4} (a).
For example, contribution of the state 11 to the $Z$-matrix is given as
\begin{eqnarray}
 q^1 (-q)^3 \left( \vec{\sigma}_4 \cdot \vec{\sigma}_3\right).
\end{eqnarray}
Including additional contributions from the states from $8$ to $15$ in Fig.~\ref{fig:Ly4} (b), 
the $Z$-matrix for PBC is given as 
\begin{eqnarray}
 \nonumber 
  Z^{\rm PBC}(4;\left\{ \vec{\sigma}_i \right\})
  &&= Z^{\rm OBC}(4;\left\{ \vec{\sigma}_i \right\})
  - q^2 \left( \vec{\sigma}_4 \cdot \vec{\sigma}_1 \right)\\
  &&+ q^3
  \left[
   \left( \vec{\sigma}_4 \cdot \vec{\sigma}_2 \right) +
   \left( \vec{\sigma}_3 \cdot \vec{\sigma}_1 \right) +
   I
  \right]\nonumber\\
  &&+ q^4
  \left[
   \left( \vec{\sigma}_4 \cdot \vec{\sigma}_1 \right) 
   \left( \vec{\sigma}_2 \cdot \vec{\sigma}_3 \right) -  
   \left( \vec{\sigma}_4 \cdot \vec{\sigma}_3 \right) \right.\nonumber\\
 &&-\left.  \left( \vec{\sigma}_3 \cdot \vec{\sigma}_2 \right) -  
  \left( \vec{\sigma}_2 \cdot \vec{\sigma}_1 \right)
  \right].
\end{eqnarray}
In principle, we can obtain an explicit form of the $Z$-matrix for larger $L_y$ using the method based on the 
graphical representations.

\begin{figure}[t]
 \begin{center}
  \includegraphics[scale=1]{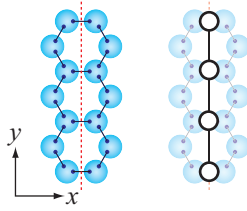}
  \caption{
  (Color online) 
  Relationship between the original VBS state on a hexagonal-lattice vertical ladder (left) and 
  a holographic spin chain (right). The circles in the right panel indicate the sites in the spin chain, 
  on which the Pauli vectors $\left\{ \vec{\sigma}_i \right\}$ in Eq. (\ref{eq:z-matrix-square}) act. 
  }
 \label{fig:z-matrix-hexagonal}
 \end{center}
\end{figure}

Next we show how to calculate the $Z$-matrices for hexagonal-lattice vertical ladders (i.e. $L_x=1$). 
The relationship between the VBS on the hexagonal-lattice vertical ladder and the holographic spin chain 
is shown in Fig.~\ref{fig:z-matrix-hexagonal}.
The only difference between the square and hexagonal vertical ladders is the definition of 
the length between sites on the holographic spin chain.
We can then calculate the $Z$-matrix for the hexagonal-lattice vertical ladder 
as follows:
\begin{eqnarray}
 \nonumber
 Z(L_y;\left\{\vec{\sigma}_i\right\})
  &=& \sum_{\left\{ i_1, \cdots, i_{2n}\right\}}
  q^n (-q)^{2l_1} \cdots (-q)^{2l_n}\\
  \label{eq:z-matrix-hexagonal}  
 &\times&  \left( \vec{\sigma}_{i_1} \cdot \vec{\sigma}_{i_2} \right)
  \cdots 
  \left( \vec{\sigma}_{i_{2n-1}} \cdot \vec{\sigma}_{i_{2n}} \right),
\end{eqnarray}
where $l_j = {\rm mod}(i_{2j}-i_{2j-1},L_y)$ and $1 \le i_1 < i_2 < \cdots < i_{2n-1} \le L_y$.
The difference between Eq.~(\ref{eq:z-matrix-square}) and Eq.~(\ref{eq:z-matrix-hexagonal}) is an exponent of $(-q)$.

After we obtain an explicit form of $Z$-matrix, we can study entanglement properties. 
Indeed, we have compared the results based on this method with those obtained by Monte Carlo method 
and have confirmed that they are completely consistent, which ensures the validity of our Monte Carlo approach. 

\section{Monte Carlo Method for Calculating Overlap Matrix}
\label{app:MC}
In this appendix, we explain our proposed Monte Carlo method used for obtaining the overlap matrix $M$.
As we mentioned in Sec. II, the VBS state can be regarded as a singlet-covering state, 
expressed by Eq.~(\ref{eq:VBS}).
In the case of bipartite lattices (square and hexagonal), 
we can avoid negative signs by a local gauge transformation:
\begin{eqnarray}
 c_i^{(1)\dagger} :=
  \begin{cases}
   a_i^\dagger  &(i \in {\cal A})\\
   b_i^\dagger  &(i \in {\cal B})
  \end{cases},
  \,\,\,
  c_i^{(2)\dagger} := 
  \begin{cases}
   b_i^\dagger  &(i \in {\cal A})\\
   -a_i^\dagger  &(i \in {\cal B})
  \end{cases},
\end{eqnarray}
where ${\cal A}$ and ${\cal B}$ denote the two sublattices. 
The VBS state can then be represented as
\begin{eqnarray}
 | \Psi \rangle = \prod_{\langle i,j \rangle} 
  ( 
   c_i^{(1)\dagger} c_j^{(1)\dagger} + c_i^{(2)\dagger} c_j^{(2)\dagger}
  )
   | {\rm vac} \rangle,
\end{eqnarray}
where $\langle i,j \rangle$ runs over all connected neighboring sites in the graph. 
If we introduce $\tau=1,2$ as the quantum number of SU(2) boson operators, 
the VBS state can be written as 
\begin{eqnarray}
 | \Psi \rangle = \sum_{\{\tau_{jk}\} } \left(
  \prod_{\langle j,k \rangle \atop j,k \in \Omega}
  c_j^{(\tau_{jk})\dagger} c_k^{(\tau_{jk})\dagger}
 \right)
 | {\rm vac} \rangle.
\end{eqnarray}
where $\tau_{jk}$ denotes the boson between $j$-th and $k$-th site on the bulk of the subsystem $\Omega$.
As a result, we can reproduce VBS state through stochastic sampling of 
boson configurations $\{\tau_{jk}\}$ of the system.

The element of overlap matrix $M$, 
is given by
\begin{eqnarray}
 M_{\{\sigma_i'\},\{\sigma_i\}} = \langle \{ \sigma_i'\}|\{\sigma_i\}\rangle,
\end{eqnarray}
where $|\{\sigma_i\} \rangle$ is the VBS state with fixed boundary $\{ \sigma_i \}$:
\begin{eqnarray}
 \nonumber
 | \{\sigma_i\} \rangle
&&= 
 \sum_{\{\tau_{jk}\}}
 \left(
   \prod_{i \in \partial \Omega} c_i^{(\sigma_i)\dagger}
  \right)
 \left(
  \prod_{\langle j,k \rangle \atop j,k \in \Omega}
  c_j^{(\tau_{jk})\dagger} c_k^{(\tau_{jk})\dagger}
 \right)
 | {\rm vac} \rangle \\
 &&= \sum_{\{\tau_{jk}\}} | \{\sigma_i\}, \{\tau_{jk}\} \rangle,
 \label{eq:sigma_state}
\end{eqnarray}
where $\partial \Omega$ denotes the boundary of $\Omega$. 
The variables $\{ \sigma_i \}, i \in \partial \Omega$ specifies the unpaired spins on the boundary.
We can further simplify the problem by considering (\ref{eq:sigma_state})
in the occupation number basis:
\begin{eqnarray}
   | \{\sigma_i\} \rangle= \sum_{\{\tau_{jk}\}} w(\{n_j^{(\alpha)}\}) | \{n_j^{(\alpha)}\} \rangle,
\end{eqnarray}
where 
$n_j^{(\alpha)} = \sum_{k \in \delta_j} \delta(\tau_{jk},\alpha)$ 
(the summation is taken over all nearest neighbor sites of $j$) is the number of bosons of 
type $\alpha$ at the site $j$. 
The degeneracy of occupation number state can be easily derived as:
\begin{eqnarray}
w(\{n_j^{(\alpha)}\}) = \prod_{j \in \Omega} \sqrt{n_j^{(1)}!n_j^{(2)}!}.
\end{eqnarray}
As a result, we obtain the element of the overlap matrix as
\begin{eqnarray}
 M_{\{\sigma_i'\},\{\sigma_i\}}
  = \sum_{\{\tau_{jk}'\},\{\tau_{jk}\}}
  W(\{n_j^{(\tau_{jk})}\}) \delta (\{n_j^{(\tau_{jk}')}\},\{n_j^{(\tau_{jk})}\}), \nonumber \\
\end{eqnarray}
with weight 
\begin{equation}
W(\{n_j^{(\tau_{jk})}\}) = w^2(n_j^{(\tau_{jk})})=\prod_{j \in \Omega} (n_j^{(1)}!n_j^{(2)}!).
\end{equation}

In the end, by generating boson states $\Sigma = (\{\tau_{jk}'\},\{\tau_{jk}\})$ 
stochastically while keeping the constraint $n_j^{(\tau' )} = n_j^{(\tau )}$ for all $j$ and $\tau$, 
we can numerically calculate the element of the overlap matrix $ M_{\{\sigma_i'\},\{\sigma_i\}}$
as accumulation number of boson configurations
with corresponding boundary condition, 
through Monte Carlo importance sampling. 
\begin{figure}[b]
 \begin{center}
  \includegraphics[scale=0.5]{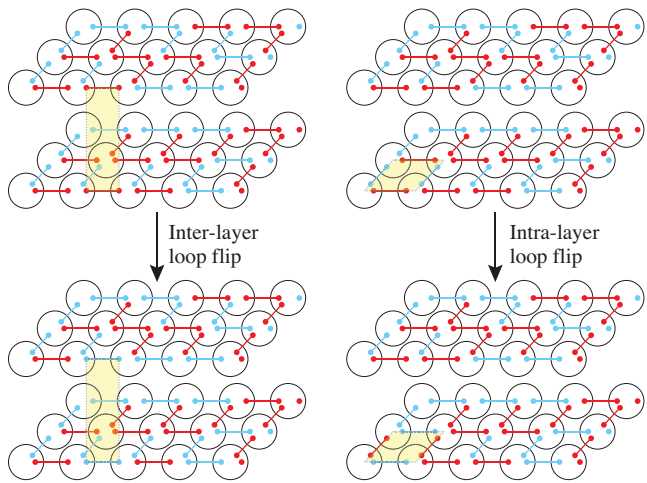}
  \caption{(Color online) Illustration of boson configurations (states) and two MC update schemes:
    (left) Inter-layer flip. (right) intra-layer flip.
    Red (dark) and blue (light) lines stand for bosons $\tau_{jk}=1,2$, respectively.
    Upper and lower layers stand for a state in sub-systems $A$ ($|\Psi^{[A]} \rangle$) and 
    its conjugate ($\langle \Psi^{[A]}|$), respectively.
    Valence bonds being flipped are indicated by a yellow (shadowed) plaquette.
    The right edge of each panel corresponds to the boundary between the subsystems $A$ and $B$, 
    and the unconnected dots represent the $\sigma$ variables.
  }
  \label{fig:mcs}
 \end{center}
\end{figure}
It is convenient to introduce graphical representation for each boson configurations
$| \{\sigma_i\},\{\tau_{jk}\}\rangle$,
as shown in Fig.~\ref{fig:mcs}. 
The red (dark) and blue (light) lines
indicate bosons with quantum number $\tau_{jk}=1$ and $2$, respectively.
Colored 
but unconnected dots on the edge correspond to boundary bosons $\{\sigma_i\}$.
Notice that we show two layers in the graphical representation, as they 
stand for the sub-system state $|\Psi^{[A]} \rangle$ and its conjugate $\langle \Psi^{[A]}|$, respectively.
Due to the fact that the full system is reflection symmetric, 
one may also consider two layers as two subsystems A and B, 
with sites (big circles) and bonds (lines) on top of each other being symmetric counterparts.

Using this graphical representation, we can easily illustrate two MC update schemes we use to
sample all possible boson configurations,
while automatically satisfying the constrain: $n_j^{(\tau')}=n_j^{(\tau)}$ for all $j$ and $\tau$.

(i){\it Inter-layer flip} (Left panel of Fig.~\ref{fig:mcs}): 
In the boson state $\Sigma$, we choose arbitrarily a valence bond connecting sites $j$ and $k$ 
such that $\tau_{jk}=\tau_{jk}'=\mu$ (colored as red/blue) is satisfied.
A trial state $\Sigma_{\rm t}$ is given as $\tau_{jk}=\tau_{jk}'=3-\mu$ (colored as blue/red) 
with all other bosons unchanged.
The update is accepted/rejected according to the ratio of probabilities $W(\Sigma_{\rm t})/W(\Sigma)$
which satisfies detailed-balance condition.
In Fig.~\ref{fig:mcs}, a yellow loop is drawn showing bosons being flipped ($\tau_{jk}=\mu\rightarrow 3-\mu$).

(ii){\it Intra-layer flip} (Right panel of Fig.~\ref{fig:mcs}):
In order to apply intra-layer flip without violating the constraints, 
we attempt to construct a ``valence bond loop'' (VBL),
or a sequence of connected singlets, 
with alternating colors ($\tau_{jk}=1\leftrightarrow 2$).
Flipping a VBL is equivalent to operating
\begin{eqnarray}
&& \prod_{\kappa=1}^{l} c_{i_{\kappa}}^{(\alpha_{\kappa})\dagger}c_{i_{\kappa+1}}^{(\alpha_{\kappa})\dagger}
c_{i_{\kappa}}^{(\beta_{\kappa})}c_{i_{\kappa+1}}^{(\beta_{\kappa})}
\nonumber \\
&& \left(
\alpha_{\kappa}=\beta_{\kappa+1}=
\begin{cases}
  \mu \, &(\kappa:{\rm odd}) \\
  \nu=3-\mu \, &(\kappa:{\rm even})
\end{cases}
\right)
\label{eqn:vbl}
\end{eqnarray}
on the VBS state (except for the weight change), where the loop is formed by $l$ sites $\{i_{\kappa}\}$ in
the order of $\kappa=1,2,...,l$,
and $\mu,\nu$ are the alternating colors along the loop.
(from red to blue and {\it vice versa}). 
In the right column of Fig.~\ref{fig:mcs}, 
we show a possible loop that consists of 4 sites and 4 valence bonds, denoted by yellow region.
In our MC simulation scheme, we construct a loop by using a random walker.
The walker starts from a randomly chosen site, as well as a randomly chosen direction.
After such a movement, we choose next direction so that the color of next valence bond alternates.
Since turning back to the previous site is in the list of candidates, 
it is always possible to choose such a direction.
When the walker comes back to the initial position, the loop is formed.
Therefore, loops can be of arbitary length, as long as color is alternating along the loop.
Meanwhile, the possibility (amplitude) decrease significantly for longer loops. 
Nevertheless, obviously, flipping all the bonds in a valence bond loop
does not affect the occupation number $n_j^{(\alpha)}$ at any site on the loop. 

Following above two updating rules, 
we can obtain efficiently the overlap matrix with high accuracy,
while satisfying the ergodicity condition.
This MC sampling approach can be applied to arbitrarily large systems. Number of sites along the bulk dimension has
limited effect on the cost of calculations. 
The computational limit is mainly related to the diagonalization of the overlap matrix, the dimension of which grows
exponentially as the number of boundary sites increases.



\begin{thebibliography}{99}
\bibitem{Amico_review}
L. Amico, R. Fazio, A. Osterloh, and V. Vedral, Rev. Mod. Phys. \textbf{80}, 517 (2008).  

\bibitem{Eisert_review}
J. Eisert, M. Cramer, and M. B. Plenio, Rev. Mod. Phys. \textbf{82}, 277 (2010). 

\bibitem{Li_Haldane}
H. Li and F. D. M. Haldane, Phys. Rev. Lett. \textbf{101}, 010504 (2008).

\bibitem{Lauchli_ES}
A. M. Lauchli, E. J. Bergholtz, J. Suorsa, and M. Haque, Phys. Rev. Lett. \textbf{104}, 156404 (2010). 

\bibitem{Thomale_ES}
R. Thomale, A. Sterdyniak, N. Regnault, and B. A. Bernevig, Phys. Rev. Lett. \textbf{104}, 180502 (2010). 

\bibitem{Fidkowski_ES}
L. Fidkowski, Phys. Rev. Lett. \textbf{104}, 130502 (2010).

\bibitem{Qi_Katsura_Ludwig}
X-L. Qi, H. Katsura, and A. W. W. Ludwig, arXiv:1103.5437v1 [cond-mat.mes-hall].

\bibitem{Calabrese_ES}
P. Calabrese and A. Lefevre, Phys. Rev A \textbf{78}, 032329 (2008).

\bibitem{Pollmann_ES}
F. Pollmann, A. M. Turner, E. Berg, and M. Oshikawa, Phys. Rev. B \textbf{81}, 064439 (2010).  

\bibitem{Poilblanc_ES}
D. Poilblanc, Phys. Rev. Lett. \textbf{105}, 077202 (2010). 

\bibitem{Yao_Qi_ES}
H. Yao and X-L. Qi, Phys. Rev. Lett. \textbf{105}, 080501 (2010). 

\bibitem{Thomale_spin_chain}
R. Thomale, D. P. Arovas, and B. A. Bernevig Phys. Rev. Lett. \textbf{105}, 116805 (2010).

\bibitem{AKLT}
I. Affleck, T. Kennedy, E. H. Lieb, and H. Tasaki, Phys. Rev. Lett. \textbf{59}, 799 (1987); 
I. Affleck, T. Kennedy, E. H. Lieb, and H. Tasaki, Commun. Math. Phys. \textbf{115}, 477 (1988). 

\bibitem{KLT}
T. Kennedy, E. H. Lieb, and H. Tasaki, J. Stat. Phys. \textbf{53}, 383 (1988).

\bibitem{Kirillov_Korepin}
A. N. Kirillov and V. E. Korepin, Algebra i Analiz. \textbf{1}, 47 (1989); 
[St. Petersburg Math. J. \textbf{1}, 343 (1990)].

\bibitem{Wei_Affleck}
T-C. Wei, I. Affleck, and R. Raussendorf, Phys. Rev. Lett. \textbf{106}, 070501 (2011).

\bibitem{Miyake}
A. Miyake, Ann. Phys. \textbf{326}, 1656 (2011).

\bibitem{Xu_Katsura_Korepin_Hirano}
Y. Xu, H. Katsura, T. Hirano, V. E. Korepin, 	J. Stat. Phys. \textbf{133}, 347 (2008). 

\bibitem{Korepin_Xu}
V. E. Korepin and Y. Xu, Int. J. Mod. Phys. B, \textbf{24}, 1361 (2010).

\bibitem{Arovas_Auerbach_Haldane}
D. Arovas, A. Auerbach, and F. D. M. Haldane, Phys. Rev. Lett. \textbf{60}, 531 (1988).

\bibitem{footnote:boundary}
Spins at the boundary sites have a different spin magnitude that is less than the bulk one ($S=z/2$).

\bibitem{KKKKT}
H. Katsura, N. Kawashima, A. N. Kirillov, V. E. Korepin, and S. Tanaka, J. Phys. A: Math. Theor. \textbf{43}, 255303 (2010). 

\bibitem{Fan_Korepin}
H. Fan, V. E. Korepin, and V. Roychowdhury, Phys. Rev. Lett. \textbf{93}, 227203 (2004).


\bibitem{desCloizeaux_Pearson}
J. des Cloizeaux and J. J. Pearson, Phys. Rev. \textbf{128}, 2131 (1962). 

\bibitem{Calabrese_Cardy}
P. Calabrese and J. Cardy, J. Stat. Mech. P06002 (2004).

\bibitem{Laflorencie_et_al}
N. Laflorencie, E. S. Sorensen, M-S. Chang, and Ian Affleck, Phys. Rev. Lett. \textbf{96}, 100603 (2006).

\bibitem{Hikihara_et_al}
T. Hikihara, S. Lukyanov, and A. Furusaki (private communication). 

\bibitem{Hida_alternating}
K. Hida, J. Phys. Soc. Jpn. \textbf{63}, 2514 (1994). 

\bibitem{Hirano-Hatsugai}
T. Hirano and Y. Hatsugai, J. Phys. Soc. Jpn. \textbf{76}, 074603 (2007).

\bibitem{Cirac_et_al}
J. I. Cirac, D. Poilblanc, N. Schuch, and F. Verstraete, Phys. Rev. B \textbf{83}, 245134 (2011).

\end{thebibliography}
\end{document}